\documentclass{aastex62}
\usepackage{epstopdf}
\usepackage{graphicx}
\usepackage[latin10]{inputenc}
\usepackage{multirow}
\usepackage[T1]{fontenc}

\usepackage{lineno}

\begin{document}
%%\title{Statistical analysis on the nonuniformity of poleward flux transport on the solar surface during solar cycles 21-24}
\title{The nonuniformity of poleward flux transport on the solar surface: a statistical method applied to solar cycles 21-24}
%%\linenumbers
\correspondingauthor{Zi-Fan Wang}
\email{zfwang@nao.cas.cn}

\author{Zi-Fan Wang}
\affiliation{Key Laboratory of Solar Activity, National Astronomical Observatories, Chinese Academy of Sciences, Beijing 100101, China}
\affiliation{School of Astronomy and Space Science, University of Chinese Academy of Sciences, Beijing, China}

\author[0000-0001-5002-0577]{Jie Jiang}
\affiliation{School of Space and Environment, Beihang University, Beijing, China}
\affiliation{Key Laboratory of Space Environment Monitoring and Information Processing of MIIT, Beijing, China}

\author{Jing-Xiu Wang}
\affiliation{School of Astronomy and Space Science, University of Chinese Academy of Sciences, Beijing, China}
\affiliation{Key Laboratory of Solar Activity, National Astronomical Observatories, Chinese Academy of Sciences, Beijing 100101, China}

\begin{abstract}
The poleward migration of the active regions' magnetic flux on the solar surface plays an important role in the development of the large-scale field development, especially the polar field reversal, which is a key process in the Babcock-Leighton-type solar dynamos.  The poleward flux transport is nonuniform, centered around poleward surges as suggested by previous observations.  The strong, long-lasting surges are related to activity complexes, and often result in violent polar field reversal.  However, the nonuniformity of poleward flux transport has not been evaluated quantitatively.  We propose a statistical method to analyze the poleward flux transport during solar cycles 21-24 by considering the frequency distributions of the magnetic field at latitudes of poleward surges occurrence during solar cycles.  The nonuniformity is quantified as the kurtosis statistics representing the tailedness of the distributions.  We test the method on results of surface flux transport simulations, and apply to WSO, NSO, MWO, and HMI data.  We confirm that the poleward surges are of significance during solar cycles 21-24 in general.  The kurtosis within a solar cycle is affected by different latitudes of the magnetic field and different data sources.  The southern hemisphere of cycle 24 exhibits the largest kurtosis, agreeing the super surge concept from previous work.  The significant nonuniformity of poleward flux transport originates from the nonrandomness of active regions, which favors the activity complexes origin of poleward surges.

\end{abstract}

\section{Introduction} \label{sec:intro}

The emergence of active regions (ARs) and the transport of AR flux on the solar surface are key processes in Babcock-Leighton-type solar dynamos that aim to explain the 11-year sunspot cycle \citep{1961ApJ...133..572B,1969ApJ...156....1L}.  Typically, an AR emerges with two opposite polarities that exhibit a tilt angle with respect to the east-west direction.  Since the tilt angles follow Joy's law \citep{1919ApJ....49..153H}, the following polarity of ARs is usually closer to the pole of the corresponding hemisphere, while the leading polarity of ARs is usually closer to the equator.  As the AR flux evolves on the solar surface, more flux of the following polarity is transported to the pole, contributing to the development of the polar field.  The polar field at cycle minimum is an important precursor of the strength of the next cycle, as shown by analysis of observations and dynamo models \citep{1978GeoRL...5..411S,2005GeoRL..32.1104S,2007MNRAS.381.1527J}.

Besides the individual AR's characteristics such as their emerging latitudes and tilt angles, the clustering tendency of ARs emergence is a notable characteristic during solar cycles.  The continuous emergence of closely located ARs during several Carrington rotations (CRs) is defined as a complex of activity \citep{1965ApJ...141.1502B,1983ApJ...265.1056G}, or an activity complex.  It is similar to the concept of sunspot nests proposed by \citet{1986SoPh..105..237C}, as a sequence of AR emergence in a compact solar surface area lasting for months.  Subsequent quantitative analysis of sunspot nests by \citet{1990SoPh..129..221B} suggests at least a third of sunspots are nested.  Activity complexes represent the important non-axisymmetric property of the large-scale field, as reviewed by \citet{2015LRSP...12....1V}.

As the emergence of ARs tends to be concentrated in activity complexes, the poleward migration of AR flux is also not uniform during solar cycles, but in the form of poleward surges (streamers, plumes) seen on the magnetic butterfly diagrams (longitudinally averaged surface magnetic field), as recognized by \citet{1981SoPh...74..131H}.  The poleward surges' strength, temporal width and influence to the polar field is determined by the originating ARs and the surface flux transport parameters \citep{2015ApJ...798..114S}.  Observationally, long-lasting activity complexes can be related to prominent poleward surges, accompanied by intense polar field reversal \citep{2017SSRv..210...77P,2019SoPh..294...21M}.  Especially, the polar field development of cycle 24 is majorly determined by a ``super surge'' at the southern hemisphere originating from activity complexes around the year 2014, as shown by \citet{2020ApJ...904...62W}.  Considering the nesting nature of ARs, super surges are likely to occur.  Such surges are able to greatly influence the short term development of the polar field, introducing violent variations, affecting the solar open field \citep{2010ApJ...709..301J}, fast solar wind \citep{2005Sci...308..519T}, and interplanetary field \citep{1995Sci...268.1007B}.  Since the polar field observations have been suspicious in the open flux models \citep{2017ApJ...848...70L}, it is important to examine the generation and development of these surges contributing to the polar field.  The surges provide us a direct perspective to understand and reconstruct the polar field development, according to the surface flux transport (SFT) models \citep{2014SSRv..186..491J}.  These surges are also highly important in the long term development of solar cycles as they can contribute a major part to the polar field at cycle minimum, and in turn to the development of following generation of toroidal field, especially the next solar cycle considered.

Previous studies on poleward surges involve the observations and simulations of specific surges, or the qualitative analysis of their relationship with the general nesting tendency of activity complexes.  SFT simulations of \citet{2015SoPh..290.3189Y} reveal the origin and contribution to the polar field of a poleward surge.  The surge they analyze originates from higher latitudes and has low influence to the polar field.  The super surge with large polar field contribution studied by \citet{2020ApJ...904...62W} is also examined with SFT simulations.  Properties of the super surge's configuration is described by the magnetic field profiles at middle to high latitudes, that is, latitudes higher than AR emergence and lower than polar regions.  The peaks (or dips) correspond to poleward surges in these profiles.  The magnetic field at middle latitudes is also used by \citet{2015ApJ...798..114S} to represent the surges as they seek correlations between surge properties and AR parameters, and by \citet{2021EGUGA..23.4954B} to study the occurrence and periodicity of surges.  The causal relationship between poleward surges and activity complexes has been evaluated qualitatively, by combining the magnetic butterfly diagrams, AR emergence data, and polar field evolutions \citep{2019SoPh..294...21M}.

%%The nonuniformity of poleward flux transport needs to be quantified.
The poleward surges originate from ARs, which are discrete in time and location during solar cycles.  That is, the magnetic field on activity latitudes cannot remain constant during one solar cycle, so the magnetic field on middle to high latitudes cannot be constant as well.  Hence, qualitatively speaking, the poleward flux transport is always nonuniform to a certain degree.  However, from the quantitative perspective, it is expected that the nonuniformity of poleward transport from ARs with and without nesting tendency will show difference, if there is a strong causal relationship between activity complexes and poleward surges.  It is required to know how nonuniform the poleward transport is quantitatively, and its possible implications on the originating ARs.  This leads us to propose a statistical method to analyze the properties of the magnetic field at middle to high latitudes.

In this article we present a statistical method to evaluate the quantitative nonuniformity of poleward flux transport during solar cycles 21-24 by analyzing the non-Gaussianity of the frequency distribution of magnetic field at middle to high latitudes, mainly expressed by the kurtosis statistic.  We establish the basis of the statistical method on the central limit theorem, and construct the relationship between poleward flux transport nonuniformity and AR emergence nonrandomness.  We validate the method with a series of SFT simulations.  We apply the method to solar cycles 21-24 from different sources of data, evaluating the kurtosis of different cycles, and the values of the data combined by the 4 cycles together.  We are able to show that generally the cycles exhibit positive kurtosis, implying significant occurrence of poleward surges.  Especially, the southern hemisphere of cycle 24 has the largest nonuniformity of poleward flux transport, indicating strong poleward surges.

This article is organized as follows.  In Section \ref{sec:method}, we introduce the statistical analysis method, and evaluate its effectiveness with SFT simulations.  We then apply the method to observational data and present the results in Section \ref{sec:results}.  We discuss and conclude in Section \ref{sec:outro}.

\section{Analysis method} \label{sec:method}

\subsection{The statistical method to quantify the nonuniformity of poleward flux transport} \label{subsec:stats}

%%Poleward surges are observed on the magnetic butterfly diagrams as poleward extending features that connect the active latitudes and polar regions.  If we focus on a certain medium to high latitude, that is, a latitude higher than the emerging latitudes of ARs and lower than polar regions, we can see that the magnetic field strength profile at the latitude forms several peaks (or dips) during the occurrence of surges.  The peak value and the peak width are direct observational characteristics of a certain poleward surge.  This evaluation method has already been utilized in the work of \citet{2020ApJ...904...62W}, in which the magnetic butterfly diagram is sliced on different latitudes in order to analyze the strength and shape of the studied prominent surge.  \citet{2021EGUGA..23.4954B} also used the peaks and dips of the magnetic field profile to analyze the occurrence of surges.

The basis of our statistical method is the central limit theorem.  The theorem states that, for a number of independent random variables satisfying a certain distribution, no matter what the distribution is, their addition will follow a Gaussian distribution as the number of the random variables becomes sufficiently large.  It is also usually expressed as, for a certain distribution, the average of a number of independent sampling will follow a Gaussian distribution, as the number of the sampling increases large enough.  The most widely applied form of the theorem has all the random variables following the same distribution.  However, more generalized forms of the central limit theorem do not require the distributions of the random variables to be exactly the same, but require that the sum is primely determined by a number of them, instead of some extreme few \citep[see e.g.][]{1922MathZ...15...221}.  This inspires us to consider the large-scale field development of the Sun.  If the emergence of ARs during a period of time is random and uncorrelated, which refers to the case where activity complexes are completely absent, the emergence of ARs will meet the prerequisites of the central limit theorem.

We consider the poleward flux transport generated by such random and uncorrelated ARs.  The magnetic field at a certain time $t$ at middle to high latitudes is an linear addition of the contribution from many ARs emerged before, as we assume that the surface flux transport process is described by a linear, kinematic magnetic induction equation on the 2 dimensional solar surface, which is often used in previous SFT studies \citep[e.g.][]{1985AuJPh..38..999D,1989ApJ...347..529W,1998ApJ...501..866V,2002SoPh..209..287M,2014ApJ...791....5J}.  This means that the nonlinear processes such as inflows toward activity belts \citep{2002ApJ...570..855H,2004ApJ...603..776Z,2004SoPh..224..217G,2010ApJ...717..597J,2012A&A...548A..57C} in the surface flux transport process are neglected.  The \emph{i}th AR with emergence time $t_{i}$ generates a contribution to the magnetic field denoted as $B_i\left(t-t_i\right)$ at time $t$, and the magnetic field can be expressed as $B\left(t\right)=\sum_{i}B_i\left(t_i-t\right)$.  If the ``$B_i$''s have the same distributions, the magnetic field can be approximated as $B\left(t\right)=\sum_{i}B_0\left(t_i-t\right)$.  If the emergence time ``$t_i$''s are randomly and independently distributed during the cycle, $B\left(t\right)$ can be regarded as an addition of repeatedly independently sampling $B_0$.  Hence, according to the central limit theorem, $B\left(t\right)$ should follow a Gaussian distribution.

The ARs in reality are not identical in terms of their characteristics such as emerging latitudes, size, and shape, so the magnetic fields at middle to high latitudes they generate are not exactly the same.  Generally, this does not affect the application of the central limit theorem under the condition that the difference between the magnetic field distributions is not too large, and that the variance of the sum is not dominated by a few particular distributions.  However, some ARs have extreme characteristics, for example, an AR with large size may emerge at higher latitudes than normal, close to the latitudes we evaluate.  In such cases, the magnetic field at the middle to high latitudes is primarily dominated by the particular AR, instead of the linear addition of many ARs.  Therefore, such magnetic fields can no longer be regarded as determined by a large number of ARs, which results in deviations from Gaussian distributions.  In observations, such ARs are possible to occur, for example, the high emerging latitudes ARs that generate the prominent surge studied by \citet{2015SoPh..290.3189Y}.

The Gaussianity, or non-Gaussianity of the distribution that $B\left(t\right)$ follows shows the statistical properties of the poleward flux transport, and the originating ARs.  If the distribution exhibits significant non-Gaussianity, it shows that the ARs generating the poleward flux transport are either too few to satisfy the central limit theorem, or are not randomly distributed and uncorrelated.  The activity complexes are a probable source of the introduction of nonrandom AR emergence.  The poleward surges, as the nonuniform component, can be referred to as the the source for the distribution deviating from the Gaussian distribution.  As the activity complexes generate more prominent poleward surges, the distribution will exhibit heavier than Gaussian tails on the sign of the polarity of the surges.  Therefore, if we get heavier than Gaussian tails of the magnetic field distributions from observational data, we can tell that the surges are not generated by random ARs.  In such cases, the nonuniformity of poleward flux transport can be considered statistically significant, as it represents important nonrandom AR characteristics that favor the production of poleward surges.  This can be an important statistical evidence to the aforementioned activity complexes -- poleward surges relationship.%%Therefore, heavier than Gaussian tails from observational data imply that the poleward transport is significantly nonuniform, and the poleward surges are not generated by random ARs, but by nonrandom ARs that favor the generation of surges.

%%The distribution that $B\left(t\right)$ follows shows the statistical properties of the poleward flux transport.  The poleward surges, as the nonuniform component, can be referred to as the part of the distribution away from the mode, that is, the ``tails'' of the distributions.  Gaussian tails indicate that the poleward flux transport is not significantly nonuniform, and the originating ARs are generally randomly distributed without the occurrence of long-lasting activity complexes.  On the contrary, heavier than Gaussian tails indicate that the poleward flux transport has more large, extreme values, which can be attributed to the prominent surges.  According to the central limit theorem, it is expected that the ARs generating the surges are either too few to satisfy the requirement of the central limit theorem, or are not randomly distributed and uncorrelated, which can be associated with activity complexes, as described by the generation of the super surge in \citet{2020ApJ...904...62W}.  Therefore, we can tell the properties of poleward flux transport and the associated ARs, by analyzing the statistical characteristics of the distribution of the magnetic field.

Hence, our statistical method analyzes the frequency distribution of the magnetic fields at middle to high latitudes, focusing on the non-Gaussianity of its tailedness.  The tailedness of a frequency distribution is often measured by the kurtosis statistical parameter, derived from the 4th central moment.  The kurtosis of a series of values $x_{1},...,x_{n}$ is defined as follows,
\begin{equation}\label{eq:kurtosis}
kurtosis=\frac{1}{n\sigma^{4}}\sum_{i=1}^{n}\left( x_{i}-\bar{x} \right)^{4}-3,
\end{equation}
where $n$ is the total number of the values, $\bar{x}$ is the mean of the values, and $\sigma$ is the standard deviation.   The kurtosis of a Gaussian distribution is 0, while the value of a distribution with heavier tail than Gaussian is larger than 0.  Hence, we use the kurtosis statistics to quantify the nonuniformity.  We get the middle to high latitude magnetic field on a certain hemisphere during a solar cycle from magnetic butterfly diagrams, and get the frequency distribution.  We then calculate and evaluate the kurtosis values.  In practice, the standard error of kurtosis (SEK) should be considered.  Approximately, the standard error is $\sqrt{24/n}$, where $n$ represents the sample size \citep[see e.g.][]{1987Int.Stat.Rev...55...163,2009QualQuant..43...481}.  We use $2\times SEK$ to judge the statistical significance of nonuniformity.  A kurtosis value large than $2\times SEK$ can be regarded as significantly nonuniform, and the originating ARs exhibit nonrandom characteristics.

\subsection{Evaluation of the statistical method with a surface flux transport model} \label{subsec:evaluation}

%%In order to evaluate whether the statistical method can quantitatively tell the difference between nonuniform poleward transport generated from nested ARs and generally uniform poleward transport generated from unnested, randomly emerging ARs, ,

In order to evaluate the ability of the statistical method to quantify the nonuniformity of poleward flux transport, and to justify whether the method can reveal the probable nonrandom ARs' characteristics that favor the formation of surges, we employ a series of SFT simulations with artificially designed source terms, generate the corresponding magnetic synoptic diagrams, and examine the magnetic field strength frequency distributions.  SFT models solve the magnetic induction equation on the solar surface for the radial component of the magnetic field, with a set of transport parameters including surface meridional flow, differential rotation and supergranular diffusion, as well as newly emerged ARs as source terms.  The SFT model that we utilize is from \citet{2004A&A...426.1075B}, with differential rotation profile adopted from \citet{1983ApJ...270..288S} and meridional flow from \citet{1998ApJ...501..866V}.  The supergranular diffusivity that we use is 500 km$^2$s$^{-1}$.

We design two cases of AR emergence, namely the unnested ARs case and nested ARs case.  Both cases are simulated without an initial magnetic field.  For the unnested case, we consider a series of AR emergence during a time period with a constant average number of ARs per unit time.  We consider an average of 1500 ARs during 10 years, hence the average number of AR emergence per day is 0.082.  The exact number of ARs emerged on a specific day during a simulation is determined by a Poisson distribution, so the emergence of ARs is random and uncorrelated.  The ARs have a size of 200 millionth of solar hemisphere, and are placed on $10^\circ$ latitude on either hemisphere, with a Gaussian scatter of latitude with $\sigma=2^\circ$.  The tilt angles of ARs are determined by a typical form of the Joy's law, which is $1.3\times\sqrt{\left|\lambda\right|}$ with $\lambda$ being the latitude.  The scatter of tilt angle follows a Gaussian profile, the width of which is determined by the AR area -- tilt scatter relationship given by \citet{2021A&A...653A..27J}.  We conduct 25 simulation runs for 10 years each.

In the nested case, the random and uncorrelated emergence across the entire 10 years is reduced by half, that is, averagely 750 ARs during 10 years.  Meanwhile, 5 concentrated nests with a length of 5 Carrington rotations are added to the simulations on 5 separate days randomly selected in the 10 years.  Each nest contains an average of 150 ARs, hence the total expectation of AR emergence during each simulation is also 1500.  Then we get the modified average AR emergence per day constructed by a low value during most of the simulations, and a high value during the relatively short nests.  The exact number of ARs emerged on each day during a simulation is also determined by a Poisson distribution based on the average AR emergence per day.  The parameters such as latitudes and tilt angles of ARs are determined by identical means as the unnested case.

The AR emergence in the SFT simulations are designed in order to generate a series of generally even poleward flux transport, and a series of nonuniform poleward flux transport mainly in the form of poleward surges.  In reality, the AR emergence during a solar cycle follows much more empirical rules from observations \citep[see e.g.][]{2011A&A...528A..82J}, which are possible to influence the nonuniformity of poleward flux transport.  For example, the latitudes of AR emergence during a solar cycle is different.  When ARs emerge at higher latitudes, the magnetic fluxes from them are less diffusive as they reach the middle to high latitudes we evaluates.  In this case the magnetic field is primarily contributed by fewer ARs, which is not consistent with the perquisites of the central limit theorem.  So the effects of these varied situations are not considered in the simulations for a clearer example.  In reality, the difference of emergence latitudes of ARs influences the analysis of the nonuniformity in the form of the latitudinal differences of kurtosis values, as shown in the following Section.

%%The nested case provides a modification to the average number of ARs per day.  The whole simulation length of 10 years is divided into alternating nesting and non-nesting periods.  The length of each period is determined by an exponential distribution with a characteristic length of 5 Carrington rotations.  During the nesting periods the average number of ARs per day is multiplied by 1.5, while during the non-nesting periods the average number is multiplied by 0.5.  The determination of actual AR emergence and AR parameters are the same of the unnested case.  The nested case also includes 25 simulations runs.

\begin{figure}
%%\plotone{}
\gridline{\fig{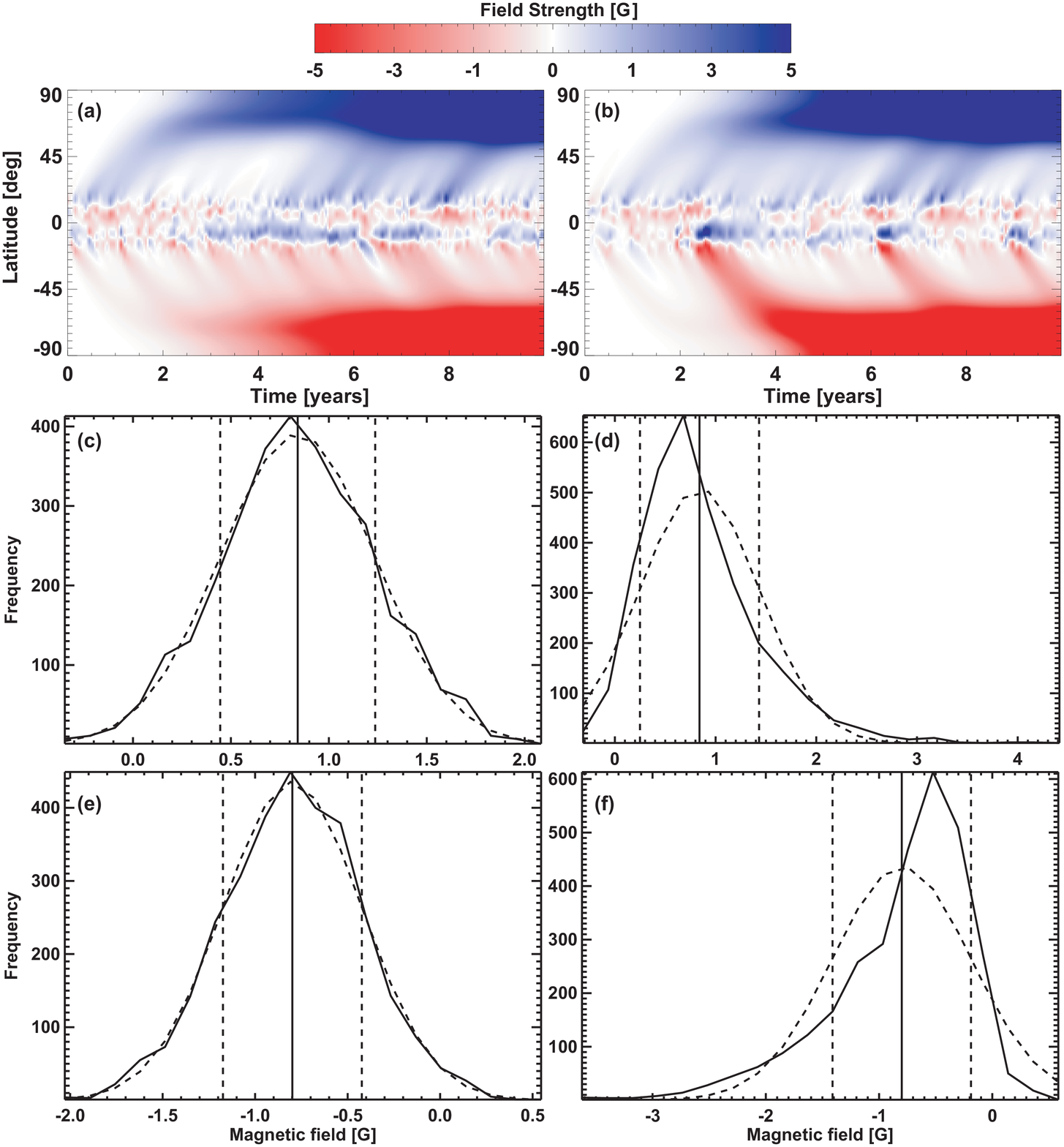}{0.8\textwidth}{}          }
\caption{Results of SFT simulations evaluating the statistical method.  Panels (a) and (b) are examples of magnetic butterfly diagrams from the unnested case and the nested case, respectively.  Panels (c), (d), (e), and (f) are magnetic field frequency distributions of the combined data of 25 runs for the two cases.  The solid curves show the distributions of the simulated data.  The dashed curves represent Gaussian profiles with identical mean and standard deviation of the distributions. The solid vertical lines mark the means, while the dashed vertical lines represent the standard deviations. Panels (c) and (e) are results for the northern and southern hemisphere for the unnested case, while panels (d) and (f) are results for the northern and southern hemisphere for the nested case.  \label{fig:sim}}
\end{figure}

The simulation results are analyzed using the method described in the previous subsection.  Each simulation contains 137 Carrington rotations, but we omit the first 14 rotations when we construct the magnetic field distributions, in order to remove the effect of the start field on the analysis.  The data of 25 runs are combined together by merging the data into one frequency distribution, as the AR emergence of each run is generated by identical means.  Hence, there are 3075 data points for each case in total.  For $n=3075$ we have $SEK=0.09$, so $2\times SEK$ is 0.18, which is used to judge the statistical significance of the non-Gaussianity of kurtosis values.  Two typical magnetic butterfly diagrams drawn from the two cases are shown in Figures \ref{fig:sim} (a) and (b).  As shown, the nested case shows clear generation of prominent poleward surges.  The unnested case shows much smoother poleward flux transport, which is close to uniform spanning across the simulation, but surges still occur occasionally if the emerged active regions are in proximity by chance.  Hence, the statistical method is needed to quantitatively distinguish the difference between the two cases.  Considering the latitudes the surges located, we choose the average of latitudes $35^\circ$ to $40^\circ$ to construct the magnetic field frequency distribution, by using 20 equal size bins across the magnetic field values.  The distributions are shown in Figures \ref{fig:sim} (c)$\sim$(f).  As shown, the distributions of the unnested case are mostly Gaussian.  The nested case, on the contrary, shows prominent non-Gaussianity in its distributions, which are not symmetric, and have heavier tails on the side of the following polarity.  Then we calculate the central moments of the magnetic field frequency distribution.  Table \ref{tab:simmoments} shows the central moments of the distributions.  As shown, both cases have the same mean value, hence the total amount of poleward flux is the same for the two cases.  The nested case has a larger standard deviation, and is more skewed against the sign of the mean, while the unnested case is generally symmetric and has close to 0 skewness, satisfying the properties of a Gaussian distribution.  The kurtosis of the unnested case is less than $2\times SEK$, while the kurtosis of the nested case is significantly greater than $2\times SEK$.  The central moments are consistent with the characteristics of the distributions shown in Figures \ref{fig:sim} (c)$\sim$(f), and the kurtosis well describes the distributions' tailedness that represents the poleward surges.   The results show that the statistical analysis based on the kurtosis effectively distinguishes the difference between the nonuniform poleward flux transport generated by nested ARs, and generally uniform poleward flux transport generated by unnested ARs.

\begin{deluxetable*}{ccccccc}[b!]
\tablecaption{Central moments of the simulated poleward magnetic field frequency distribution \label{tab:simmoments}}
\tablecolumns{6}
\tablenum{1}
\tablewidth{0pt}
\tablehead{
\colhead{Sunspot nests} &
\colhead{Latitudes} &
\colhead{Mean} &
\colhead{Standard deviation} &
\colhead{Skewness} &
\colhead{Kurtosis} &
}
\startdata
       \multirow{2}{1cm}{Unnested}   &  $-35^\circ \sim -40^\circ$  &    -0.80   &   0.38  &  -0.06 &     0.00    \\
       ~   &  $35^\circ \sim 40^\circ $ &   0.84   &   0.40   & -0.02  &    -0.12         \\
       \multirow{2}{1cm}{Nested}   &  $-35^\circ \sim -40^\circ$  &    -0.80   &   0.61  &   -1.19  &    1.76    \\
       ~   &  $35^\circ \sim 40^\circ $ &  0.84    &  0.59  &   1.20   &   2.70       \\
\enddata
\end{deluxetable*}

%%We also increase the credibility of our analysis by comparing WSO results to other observational sources.  We consider the data from National Solar Observatory (NSO) covering cycles 21-24, and the data from Mount Wilson Observatory (MWO) covering cycles 21-23.  The synoptic maps of these sources are used to generate corresponding magnetic butterfly diagrams, and the magnetic field on medium to high latitudes on both hemispheres is used to proceed the aforementioned analysis.  The statistical analysis results are presented in the following section.

\section{Analysis of observational data}\label{sec:results}

\begin{figure}
%%\plotone{}
\gridline{\fig{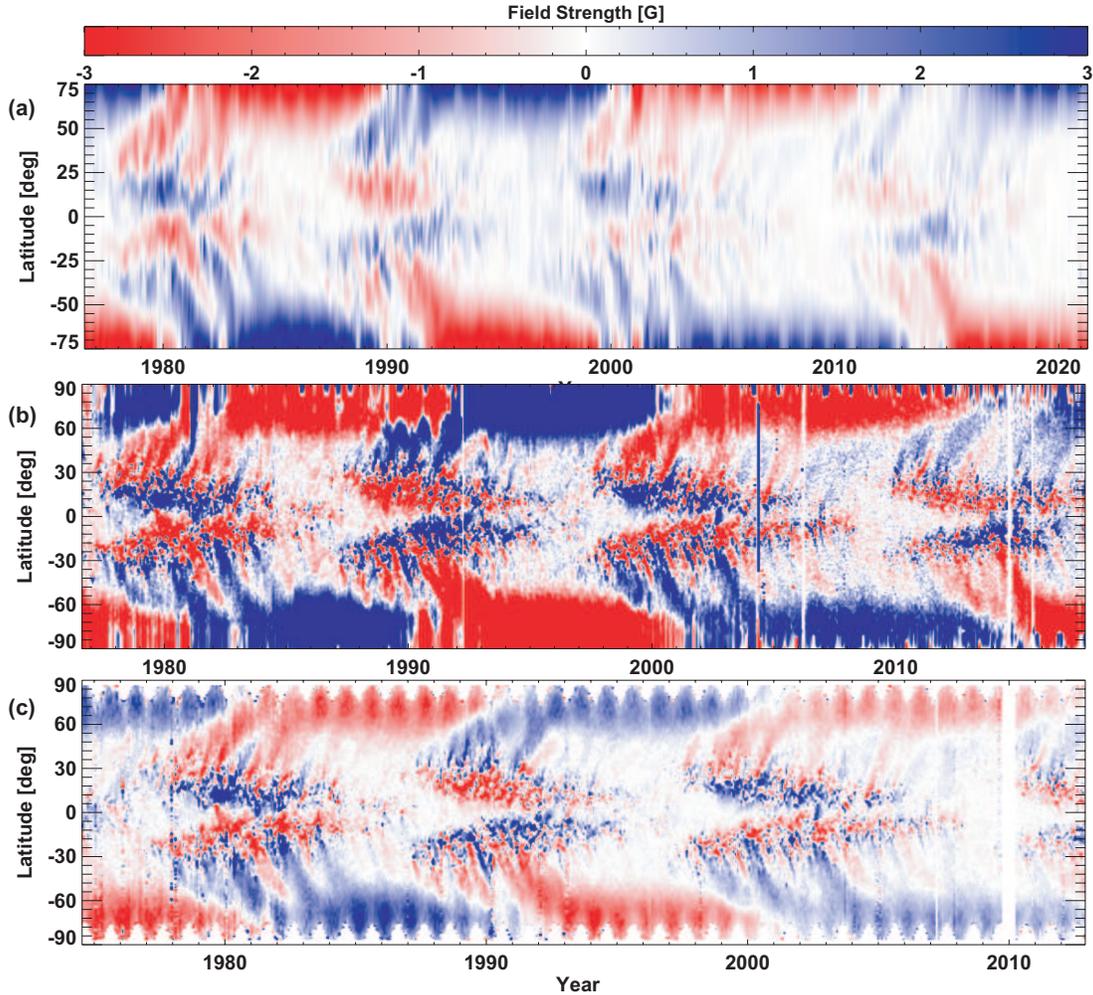}{0.8\textwidth}{}          }
\caption{Magnetic butterfly diagrams generated from longitudinally averaged synoptic maps.  Panels (a), (b), and (c) are generated from WSO, NSO, and MWO data, respectively.  The diagrams are equally spaced in latitude.\label{fig:magbfly}}
\end{figure}

In this section we quantify the nonuniformity of solar cycles 21-24 using the statistical method established in the previous section.  We employ the surface line-of-sight magnetic field synoptic maps from Wilcox Solar Observatory (WSO), and surface radial magnetic field synoptic maps from National Solar Observatory (NSO) and Mount Wilson Observatory (MWO).  The WSO data and NSO data cover solar cycles 21-24, while the MWO data covers solar cycles 21-23.  We also include the more recent results from the synoptic maps of Helioseismic and Magnetic Imager of Solar Dynamical Observatories (SDO/HMI) for solar cycle 24 for better quality.  The magnetic butterfly diagrams generated from WSO, NSO, and MWO data are displayed in Figure \ref{fig:magbfly}.  The poleward surges can be clearly seen on all diagrams.  However, different sources of data have different resolutions and hence show different levels of details, which can be directly expressed in the magnetic butterfly diagrams.  Generally, the WSO data shows the least details of the surges, while the MWO data shows the most.  This may lead to differences in the results of different sources in the following analysis.

We analyze the magnetic field strength frequency distribution at middle to high latitudes.  The WSO data is composed of 30 data points equally spaced in sine latitude between $-75^\circ$ and $75^\circ$, so we choose the latitudes $34^\circ$, $39^\circ$, and $44^\circ$ for both hemispheres from the original data points.  The NSO/KPVT data and the NSO/SOLIS data have 180 and 900 data points equally spaced
in sine latitudes, respectively.  The MWO data has 512 points equally spaced in latitudes.  The HMI data has 1440 points equally spaced in sine latitudes.   In order to compare the results of NSO, MWO, and HMI with the results of WSO, we choose to use the average of several close latitudes when we deal with NSO, MWO, and HMI data.  The latitudes $35^{\circ}$-$40^{\circ}$ and $40^{\circ}$-$45^{\circ}$ are used for NSO, MWO, and HMI.  Then we construct the frequency distribution and calculate the kurtosis values at these latitudes.

The kurtosis values for each cycle, hemisphere, and latitude considered from the 4 data sources are shown in Tables \ref{tab:wsomoments}, \ref{tab:nsomoments}, \ref{tab:mwomoments}, and \ref{tab:hmimoments}.  The values are positive for many cycles and latitudes, and are subject to changes for different latitudes, cycles, and data sources.  Averagely, each solar cycle lasts for 11 years, that is, approximately 149 Carrington rotations. For $n=149$, $SEK=0.4$.  Hence the $2\times SEK$ criterion should be 0.8, which will be used in the analysis to judge statistical significance.

%%The 4 central moments, namely the mean, standard deviation, skewness, and kurtosis for each cycle, hemisphere, and latitude considered from the 3 data sources are shown in Tables \ref{tab:wsomoments}, \ref{tab:nsomoments}, and \ref{tab:mwomoments}.  The mean of the distributions is a representation of the total amount of flux transported toward the poles.  For the two hemispheres of one cycle, the amounts of poleward flux are roughly similar, but some exceptions still exist.  The standard deviation varies for different cycles and latitudes.  The skewness is generally of the same sign as the mean, indicating a heavier tail of the distribution on the side of following polarity, as more surges are of the following polarity.  The kurtosis, as the key of our analysis, is positive for many cycles and latitudes, and is subject to changes in different cases.  Considering the length of solar cycles, the standard error of kurtosis for one cycle is approximately 0.4.  Hence the $2\times$ standard deviation criterion should be 0.8, which will be used in the analysis to judge statistical significance.  %%The variation of kurtosis is probably due to cycle, hemispheric, and latitudinal differences, as well as the limited data points during one solar cycle.  Considering the length of solar cycles, the standard error of kurtosis for one cycle is approximately 0.4, which will be used in the following kurtosis analysis.

The kurtosis values show latitudinal variations within a certain hemisphere and cycle, which shows statistical significance in general.  We consider the kurtosis value variations statistically significant if the variations are larger than $2\times SEK=0.8$.  Among the 22 cases considering all hemispheres, cycles, and data sources, about 68\% show a statistically significant decrease in kurtosis values as the latitude increases.  The general decrease of kurtosis is consistent with the considerations of our statistical method, as poleward surges tend to be more diffusive as they migrate to higher latitudes, which makes the overall poleward transport smoother.  In this case the distribution has less values in the tails, so the kurtosis values would be smaller.  Meanwhile, since the magnetic field at higher latitudes can be associated with more ARs as a result of AR flux being more diffusive on reaching higher latitudes, they are more consistent with the requirement of the central limit theorem.  Hence, they tend to follow a distribution closer to a Gaussian distribution.  However, we also note that there are exceptions to the general trend, as there are about 14\% of the cases show a statistically significant increase in kurtosis values, such as the southern hemisphere of cycle 23 for NSO and MWO.  The originating latitudes of the surges are different, which is a possible cause of kurtosis latitudinal variations, especially the variations opposite to the general trend.

The kurtosis values also show differences for different data sources.  Considering the $2\times SEK$ criterion for judging non-Gaussianity, some solar cycles have an agreement in terms of non-Gaussianity of poleward flux transport distribution, while some cycles do not have an agreement.  For solar cycles 21 and 23, the WSO data source does not show significant non-Gaussianity, while both the NSO and MWO data sources show significant non-Gaussianity at some hemispheres and latitudes.  This can be possibly explained by the fact that the different data sources have different resolutions, and show different levels of details of the surges in the magnetic butterfly diagrams. As described above in Figure \ref{fig:magbfly}, the details of the surges are less displayed in the diagram of WSO than in NSO and MWO, which is possible to affect the results of the kurtosis analysis.  More details of the surge generally mean more uneven structures in the poleward flux transport, increasing overall nonuniformity.  This interpretation agrees with the result that the largest kurtosis value 6.33 is obtained from HMI data on the southern hemisphere of cycle 24, as the HMI data shows the most detailed structure of cycle 24.  The difference between the HMI results and other results of the southern hemisphere is larger than $2\times SEK=0.8$, indicating a significant data quality difference.

The kurtosis values represent the characteristics of poleward flux transport in different hemispheres and solar cycles.  From the results with better agreement among the data sources, we can see that the northern hemisphere of solar cycle 22 and the southern hemisphere of cycle 24 have kurtosis values that are larger than the value of a Gaussian distribution of statistical significance.  This indicates that the poleward flux transport during these cycles on the certain hemisphere is more uneven than what randomly generated ARs could produce, so the associated ARs emergence should not be random. The results of the southern hemisphere of cycle 24 have the largest kurtosis values, and are highly in agreement among the results of different latitudes and data sources, which agrees with the result of \citet{2020ApJ...904...62W} that the super surge considered has a profound and long-lasting effect on the development of the solar cycle.  The northern hemisphere, on the contrary, does not show non-Gaussianity of statistical significance.  The results of the HMI data are consistent with the interpretation of cycle 24, as the kurtosis on the southern hemisphere is essentially large.  The northern hemisphere has significantly lower kurtosis values.  However, since the HMI data source tends to produce larger kurtosis values than other data sources in general, the kurtosis on $35^\circ \sim 40^\circ$ still exceeds $2\times SEK=0.8$.  Still, the super surge occurrence on the southern hemisphere is strongly favored by HMI results.

\begin{deluxetable*}{cccccc}[b!]
\tablecaption{Kurtosis of the frequency distribution of poleward magnetic field based on the WSO data for cycles 21-24 \label{tab:wsomoments}}
\tablecolumns{5}
\tablenum{2}
\tablewidth{0pt}
\tablehead{
\colhead{Latitudes} &
\colhead{Cycle 21} &
\colhead{Cycle 22} &
\colhead{Cycle 23} &
\colhead{Cycle 24} &
}
\startdata
  $-34^\circ$  &     0.98 &    1.56 &    1.15  &    3.70    \\
  $-39^\circ$ &     -0.32 &    0.48   &    0.19  &    3.26         \\
  $-44^\circ$ &    -0.28  &    -0.13  &    0.50   &    3.60       \\
  $34^\circ$ &    -0.43  &    2.80    &    0.39   &    0.17     \\
  $39^\circ$ &   0.44   &    2.44   &   -0.38   &   -0.87      \\
  $44^\circ$ &    0.79   &    1.06  &    -0.44  &    -0.82        \\
\enddata
\end{deluxetable*}

\begin{deluxetable*}{cccccc}[b!]
\tablecaption{Kurtosis of the frequency distribution of poleward magnetic field based on the NSO data for cycles 21-24 \label{tab:nsomoments}}
\tablecolumns{5}
\tablenum{3}
\tablewidth{0pt}
\tablehead{
\colhead{Latitudes} &
\colhead{Cycle 21} &
\colhead{Cycle 22} &
\colhead{Cycle 23} &
\colhead{Cycle 24} &
}
\startdata
    $-35^\circ \sim -40^\circ$  &      -0.13  &    1.52  &    0.07 &    3.55    \\
    $-40^\circ \sim -45^\circ$ &     -0.11  &    -0.37   &    1.23  &    2.67     \\
    $35^\circ \sim 40^\circ$ &     2.27    &    2.13   &    1.77  &    0.46       \\
    $40^\circ \sim 45^\circ$ &      1.09     &    0.73   &   1.06  &   -0.62     \\
\enddata
\end{deluxetable*}

\begin{deluxetable*}{ccccc}[b!]
\tablecaption{Kurtosis of the frequency distribution of poleward magnetic field based on the MWO data for cycles 21-23 \label{tab:mwomoments}}
\tablecolumns{4}
\tablenum{4}
\tablewidth{0pt}
\tablehead{
\colhead{Latitudes} &
\colhead{Cycle 21} &
\colhead{Cycle 22} &
\colhead{Cycle 23} &
}
\startdata
         $-35^\circ \sim -40^\circ$  &      1.19  &    2.86 &    0.30 \\
         $-40^\circ \sim -45^\circ$ &   -0.04   &    0.45   &    1.15    \\
         $35^\circ \sim 40^\circ$ &     3.88  &    1.60   &    1.28        \\
         $40^\circ \sim 45^\circ$ &     1.31  &    1.26    &   0.43     \\
\enddata
\end{deluxetable*}

\begin{deluxetable*}{ccccc}[b!]
\tablecaption{Kurtosis of the frequency distribution of poleward magnetic field based on the HMI data for cycle 24
\label{tab:hmimoments}}
\tablecolumns{2}
\tablenum{5}
\tablewidth{0pt}
\tablehead{
\colhead{Latitudes} &
\colhead{Cycle 24} &
}
\startdata
         $-35^\circ \sim -40^\circ$  &      6.33  \\
         $-40^\circ \sim -45^\circ$ &   4.22      \\
         $35^\circ \sim 40^\circ$ &     1.32        \\
         $40^\circ \sim 45^\circ$ &     -0.30      \\
\enddata
\end{deluxetable*}

We further apply the analysis method to the data sets combining the data from different solar cycles.  As shown above, the results of an individual cycle show nonuniformity at some cycles, while some other cycles do not have nonuniformity with statistical significance.  In order to answer whether the poleward flux transport is nonuniform in general terms, combining the data from different cycles is needed.  Combining the data also decreases the standard error of kurtosis as the sample size is increased.  The data of an individual cycle suffers from a relatively high standard error, which limits the credibility.  The general results from the combined data can be more credible when judging the statistical significance of kurtosis.  We consider that the total amount of poleward transported flux is different for different cycles, hence the absolute values of ``tails'' of the frequency distributions should be different.  Therefore, we normalize the data from different cycles by dividing the data by the mean of the corresponding cycle and hemisphere.  Then the frequency distribution is constructed for each data source and the different orders of central moments are obtained.

%%A practical approach to reduce the standard error and increase credibility of statistical results is to increase the sample size.  Hence, we combine the data of different solar cycles into one data set for each data source considered.  Although the cycle-specific results are no longer obtainable in this case, we get more precise results describing the general level of significance regarding the poleward surges during all solar cycles considered.  We consider that the total amount of poleward flux transport is different for different cycles, hence the absolute value of ``tails'' of the frequency distributions should be different.  Therefore, we normalize the data from different cycles by dividing the data by the mean of each cycle and hemisphere.  Then the frequency distribution is constructed for each data source and the different orders of central moments are obtained.

\begin{figure}
%%\plotone{}
\gridline{\fig{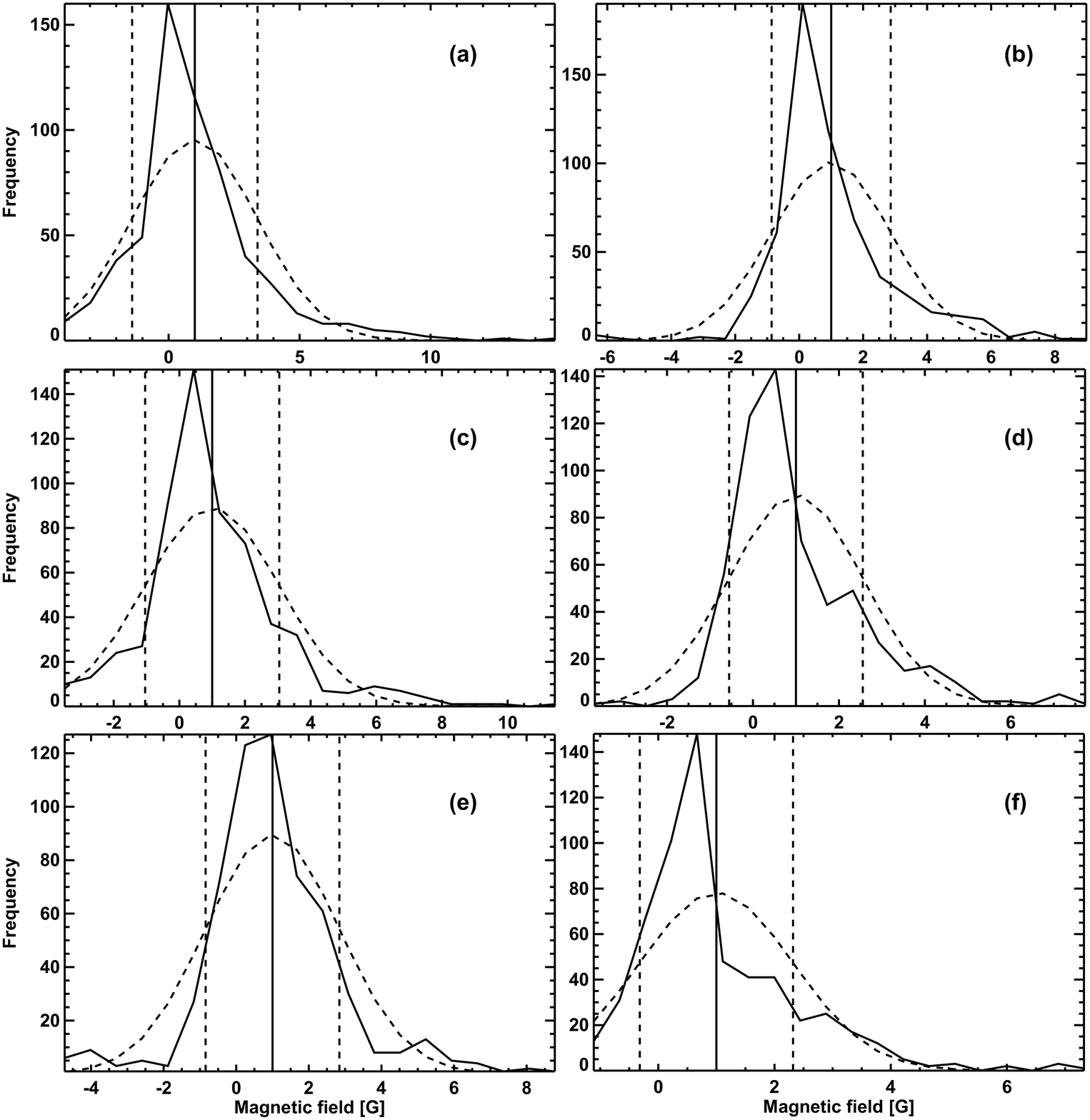}{0.8\textwidth}{}          }
\caption{Magnetic field strength distributions for the WSO data during cycles 21-24, on different latitudes.  The solid curves show the distributions of the observed data.  The dashed curves represent Gaussian profiles with identical means and standard deviations of the corresponding distributions.  The solid vertical lines mark the means, while the dashed vertical lines represent the standard deviations.  Panels (a) to (f) refer to latitudes $35^\circ$, $-35^\circ$, $40^\circ$, $-40^\circ$, $45^\circ$, and $-45^\circ$, respectively.  \label{fig:wsoresults}}
\end{figure}

We present the frequency distributions of the poleward magnetic field based on the WSO data as a typical example.  The distributions for different latitudes of the WSO data are displayed in Figure \ref{fig:wsoresults}.  A Gaussian profile with identical mean and standard deviation is also plotted in each panel, to show the difference between the distributions and a Gaussian distribution.  The distributions show significant non-Gaussianity, as they are much different from the Gaussian profiles.  They are constructed by a narrower peak, and a heavier tail at the following polarity side, much similar to the frequency distribution of the nested case described in Subsection \ref{subsec:evaluation}.  Note that since we divide the data of an individual cycle by the mean of the corresponding cycle and hemisphere, the mean of the data set for cycles 21-24 is 1.  The heavy tails imply that the contribution of poleward surges to the poleward flux transport during cycles 21-24 is higher than the poleward magnetic field satisfying a Gaussian distribution in general, indicating a non-random AR emergence during the cycles.

The central moments for the combined data of all three data sources are shown in Table \ref{tab:allmoments}.  As shown, the skewness and kurtosis are all positive, which is consistent with Figure \ref{fig:wsoresults} that shows heavier tails especially on the positive side.  For the data combining cycles 21-24 (21-23 for MWO), the $SEK$ value is 0.2 (0.20 for 4 cycles, and 0.23 for 3 cycles).  Compared to $2\times SEK=0.4$, the kurtosis values for all cases listed show the significant non-Gaussian property.  Hence, the data sources agree that the poleward surges during the cycles considered have statistical significance in terms of kurtosis.  Meanwhile, there exist latitudinal and hemispheric differences with statistical significance compared to $2\times SEK$, for all data sources considered.  The results suggest that the poleward surges, as the nonuniform component, play an important role in the poleward transport of AR flux.  The surges are not generated by some random ARs that happen to emerge in proximity by chance, but generated by the AR emergence that is not random and uncorrelated during the solar cycles.   %%The poleward surges are not simply generated randomly, but have higher rate of occurrence and strength.

\begin{deluxetable*}{cccccc}[b!]
\tablecaption{Central moments of the frequency distributions of poleward magnetic field for WSO, NSO, and MWO \label{tab:allmoments}}
\tablecolumns{5}
\tablenum{6}
\tablewidth{0pt}
\tablehead{
\colhead{Observation source} &
\colhead{Latitudes} &
%\colhead{Mean} &
\colhead{Standard deviation} &
\colhead{Skewness} &
\colhead{Kurtosis} &
}
\startdata
       \multirow{6}{1cm}{WSO}   &  $-34^\circ$  &      1.86  &  0.83 &    3.14    \\
       ~   &  $-39^\circ$ &     1.56   & 1.22  &    2.42         \\
       ~   &  $-44^\circ$ &     1.32   & 1.42  &    2.88         \\
       ~   &  $34^\circ$ &      2.39   & 1.33  &    3.94         \\
~   &  $39^\circ$ &     2.04   & 0.96  &    2.48         \\
~   &  $44^\circ$ &      1.84   & 0.35  &    2.47         \\
       \hline
       \multirow{4}{1cm}{NSO}   &  $-35^\circ \sim -40^\circ$  &     2.31  &  0.62 &    1.47    \\
       ~   &  $-40^\circ \sim -45^\circ$ &     1.78  & 0.89  &    1.30         \\
       ~   &  $35^\circ \sim 40^\circ$ &      2.24   & 0.85  &    4.39       \\
~   &  $40^\circ \sim 45^\circ$ &     1.96   & 0.64  &    2.35         \\
%       ~   &  $-45^\circ \sim -50^\circ$ &     0.31   & -0.75  &    -0.13        \\
%       ~   &  $45^\circ \sim 50^\circ$ &      0.26   & 0.89  &    1.06         \\
       \hline
       \multirow{4}{1cm}{MWO}   &  $-35^\circ \sim -40^\circ$  &     2.02  &  0.36 &    1.82    \\
       ~   &  $-40^\circ \sim -45^\circ$ &    1.50  & 0.75  &    1.03         \\
       ~   &  $35^\circ \sim 40^\circ$ &      2.40   & 1.41  &    3.85       \\
~   &  $40^\circ \sim 45^\circ$ &     1.94   & 1.31  &    1.65         \\
 %      ~   &  $-44^\circ$ &     0.15   & 1.13  &    0.50        \\
%       ~   &  $44^\circ$ &      0.11   & -0.20  &    -0.44         \\
\enddata
\tablecomments{As the combination of the data for different cycles involving dividing the data by the mean of the cycle, the 1st order central moments, i.e., the means for the combined data should be 1 for all cases, hence they are not listed here.}
\end{deluxetable*}

\section{Discussion and conclusion}\label{sec:outro}

In the article, we quantitatively describe and analyze the nonuniformity of poleward flux transport, which manifests as poleward surges on magnetic butterfly diagrams, by a statistical approach we propose.  We investigate the magnetic field frequency distribution at middle to high latitudes.  The 4th central moment kurtosis describing the tailedness of the distribution is a representation of the magnetic fields with large, extreme values of the poleward flux transport, hence implying the occurrence of strong poleward surges.  For randomly and independently emerging ARs, the resulting poleward flux transport should follow a Gaussian distribution as a result of the central limit theorem.  Therefore, a distribution with a larger kurtosis than that of the Gaussian distribution indicates that the poleward flux transport is significantly nonuniform, characterized by prominent surges probably generated by ARs with nesting tendencies.  The statistical method is examined by a series of SFT simulations, confirming its ability to quantify the nonuniformity of poleward flux transport, and to distinguish poleward flux transport generated by ARs with and without nesting tendency.

We apply the statistical method to the surface magnetic field synoptic map data of WSO, NSO, MWO, and HMI.  Generally, the poleward flux transport during the solar cycles 21-24 exhibits considerable nonuniformity.  The kurtosis for the data sets combining all cycles is higher than $2\times$ standard error.  For the kurtosis of an individual cycle, latitudinal, hemispheric, and data source differences are observed.  The kurtosis tends to decrease at higher latitudes, as the poleward surges become more diffusive at higher latitudes.  The kurtosis values are subject to differences in different data sources, as they tend to show different levels of details of the surges in their magnetic butterfly diagrams.  The characteristic of the cycles and hemispheres can be shown from the results.  Concerning the agreement of data sources, the northern hemisphere of cycle 22 has significant poleward surges from the statistical view.  The southern hemisphere of cycle 24 has the highest kurtosis values among the cycles, while the northern hemisphere does not have non-Gaussian kurtosis of statistical significance.  This can be attributed to the super surge on the southern hemisphere during cycle 24 studied by \citet{2020ApJ...904...62W}.

The significant nonuniformity of poleward flux transport during solar cycles 21-24 indicates that the generation of the poleward surges is not a totally random process.  The ARs generating the poleward surges deviate from the requirements of the central limit theorem.  They may not be randomly, independently generated, which agrees with the concept that the long-lasting activity complexes introduce poleward surges \citep{2017SSRv..210...77P,2019SoPh..294...21M}.  This means that the poleward surges during solar cycles we consider are not generated by chance.  Instead, they are likely to be generated as a result of the nesting tendency of ARs.%%We conclude that the AR emergence during solar cycles is not random in general, but further differentiating the case of activity complexes and the case of high latitude emergence is needed in the future.

As we apply the method to obtain the results of an individual cycle, we meet the limitations of the relatively large standard error of the kurtosis as a result of limited data points during one solar cycle.  Meanwhile, the latitudinal variations and different data sources complicate the method and the analysis of the results.  Further utilization of the method needs more suited latitudes to express the characteristics of surges, and more consistent synoptic magnetic field data with good reliability.  Especially, the latitudinal variations of magnetic field distributions are related to the diffusion of magnetic fluxes as they migrate poleward, as well as the different emerging latitudes of the originating ARs of poleward flux transport.  The possible AR emergence at relatively higher latitudes is also able to introduce non-Gaussianity to the distribution.  These factors are complicated components of the analysis that should be considered as we improve the method.

The statistical view in this work is important in the analysis of poleward surges, and we expect it to be further utilized in analyzing the solar large-scale field development.  The statistical view paves the way of identifying properties of the large-scale field.  This removes the ambiguity when we discuss the relatively strong or weak magnetic phenomena of the large-scale field.  Meanwhile, there are other cases in the development of the large-scale field where the magnetic field is contributed by a number of sources.  For example, the polar field at a certain time is contributed by a large number of ARs emerged before.  Hence, it is possible to further utilize the central limit theorem in the analysis of the polar field evolution, and other perspectives of the large scale field.  However, we state that the current method utilizes 1 dimensional data that only focus on the nonuniformity, while other details of the poleward flux transport are neglected.  This is because that our focus in this article concerns primarily the features on the magnetic butterfly diagram, which does not include longitudinal information. It is expected that more advanced statistical method will be utilized in analyzing the solar large-scale field, especially the 2 dimensional methods, in order to obtain more results.  For example, the mixture of Kent distributions on 2 dimensional spherical surface \citep{1982J.R.Statist.Soc.B...44..71} can represent the distribution and evolution of the surface magnetic features with more detailed results, and is expected to be utilized in the future.%%    Especially, the analysis of Gaussianity based on the central limit theorem can also be applied to other cases where the magnetic field is contributed by a number of sources, for example, the polar field.  Such studies may reveal more characteristics of ARs, and their influence to the solar large scale field. %% The statistical perspective also proposes a quantified standard for phenomena of the large scale field.  In our study, the poleward surges can be quantified as the tails of the frequency distribution, that is, the magnetic field values that deviate from the peak of the Gaussian component.  This removes the ambiguity when we discuss the relatively strong or weak magnetic phenomena of the large scale field.

\acknowledgments

Wilcox Solar Observatory data was obtained via the web site http://wso.stanford.edu.  The Wilcox Solar Observatory is currently supported by NASA.  KPVT and SOLIS data are obtained by the NSO Integrated Synoptic Program, managed by the National Solar Observatory, which is operated by the Association of Universities for Research in Astronomy (AURA), Inc. under a cooperative agreement with the National Science Foundation.  The MWO data are from the synoptic program at the 150-Foot Solar Tower of the Mt. Wilson Observatory.  The Mt. Wilson 150-Foot Solar Tower is operated by UCLA, with funding from NASA, ONR and NSF, under agreement with the Mt. Wilson Institute.  The SDO/HMI data are courtesy of NASA and the SDO/HMI team.  This research was supported by the Strategic
Priority Program of Chinese Academy of Sciences, Grant No. XDB41000000, the National Natural Science Foundation of China through grant Nos. 11873023, 11873059, and 12173005, Key Research Program of Frontier Sciences of CAS through grant No. ZDBS-LY-SLH013, and Yunnan Academician Workstation of Wang Jingxiu (No. 202005AF150025).

\end{document}